\documentclass[10pt,twocolumn,superscriptaddress,amssymb,amsmath,aps,prl]{revtex4-1}
\usepackage{graphicx}

\usepackage[normalem]{ulem}

\usepackage{color}

\renewcommand{\figurename}{\textbf{Fig.}}

\renewcommand{\tablename}{\textbf{Table}}

\renewcommand{\thetable}{\arabic{table}}
\makeatletter
\def\fnum@figure{\figurename\nobreakspace\textbf{\thefigure}}
\def\fnum@table{\tablename\nobreakspace\textbf{\thetable}}
\makeatother

\begin{document}

\setcitestyle{super}

\title{Negative radiation pressure in metamaterials explained by light-driven atomic mass density rarefication waves}
\date{April 26, 2022}
\author{Mikko Partanen}
\affiliation{Photonics Group, Department of Electronics and Nanoengineering, Aalto University, P.O. Box 13500, 00076 Aalto, Finland}
\author{Jukka Tulkki}
\affiliation{Engineered Nanosystems Group, School of Science, Aalto University, P.O. Box 12200, 00076 Aalto, Finland}

\begin{abstract}
The momentum and radiation pressure of light in negative-index metamaterials (NIMs) are commonly expected to reverse their direction from what is observed for normal materials. The negative refraction and inverse Doppler effect of light in NIMs have been experimentally observed, but the equally surprising phenomenon, the negative radiation pressure of light, still lacks experimental verification. We show by simulating the exact position- and time-dependent field-material dynamics in NIMs that the momentum and radiation pressure of light in NIMs can be either positive or negative depending on their subwavelength structure. In NIMs exhibiting negative radiation pressure, the negative total momentum of light is caused by the sum of the positive momentum of the electromagnetic field and the negative momentum of the material. The negative momentum of the material results from the optical force density, which drives atoms backward and reduces the local density of atoms at the site of the light field. In contrast to earlier works, light in NIMs exhibiting negative radiation pressure has both negative total momentum and energy. For the experimental discovery of the negative radiation pressure, one must carefully design the NIM structure and record the joint total pressure of the field and material momentum components.
\end{abstract}

\maketitle

\begin{figure}
\centering
\includegraphics[width=\columnwidth]{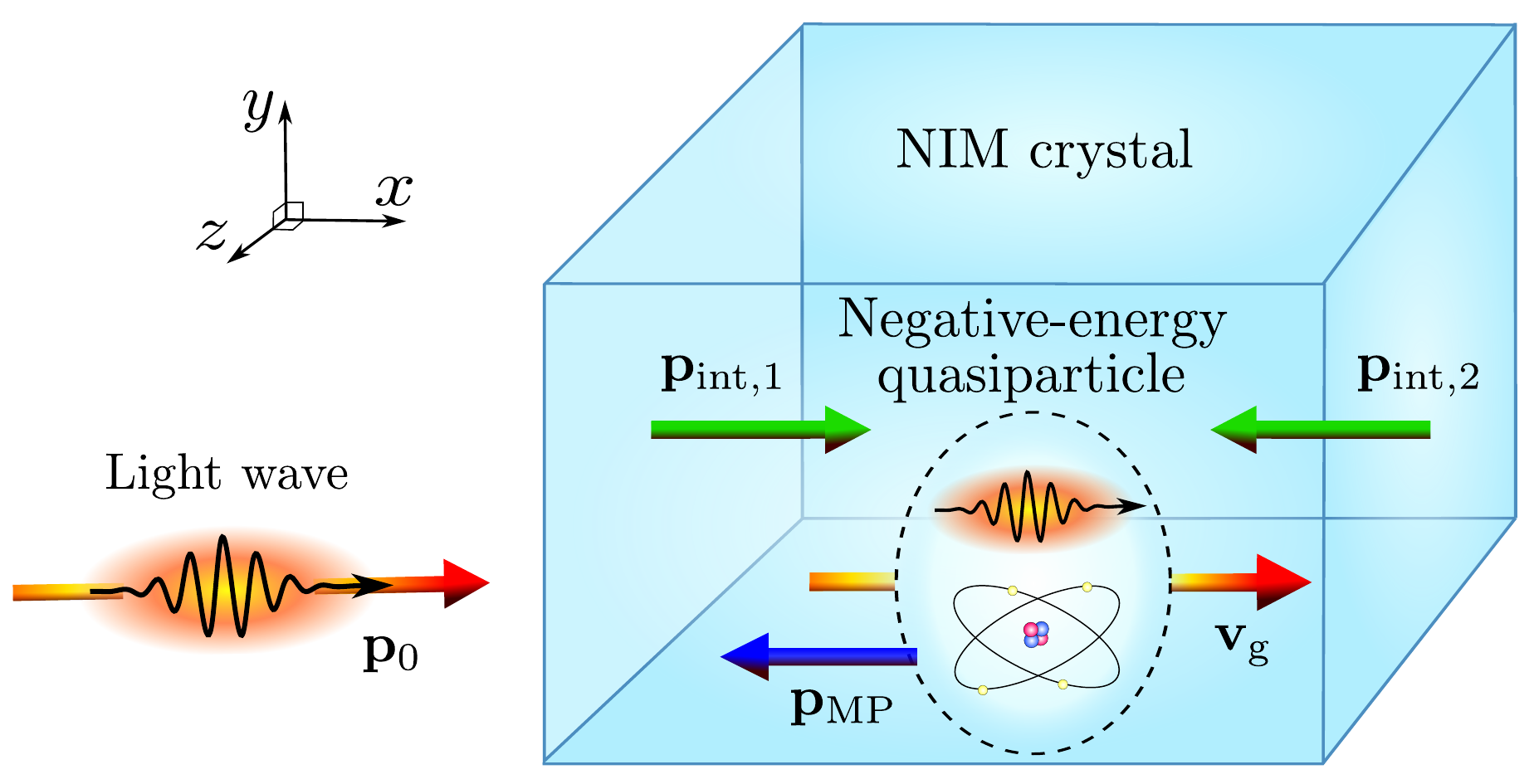}
\caption{\label{fig:illustration}
\textbf{Schematic illustration of the coupled state of light in a NIM crystal exhibiting negative radiation pressure.} The optical field is incident from the left. Inside the NIM crystal, it forms a coupled quasiparticle state with atomic displacements driven by the optical force density. When the field enters or exits the NIM crystal, the interfaces of the crystal take recoil momenta due to the momentum change of the EM wave in vacuum and the coupled state in the NIM crystal. In contrast to normal materials, the energy of the coupled state is negative, and its momentum is directed backward. The directions of the interface momenta are also reversed from what is observed for normal materials.}
\end{figure}

Negative-index metamaterials (NIMs) exhibit a number of extraordinary optical phenomena ranging from the inverse Doppler effect to beating the diffraction limit with superlenses \cite{ChenJi2011,Pendry2000,Shelby2001,Shalaev2007,Suzuki2018,Valentine2008,Smith2004,Pendry2004,Kadic2019}. The possibility of NIMs was originally hypothesized by Veselago in 1960s \cite{Veselago1968}. Among the astonishing phenomena predicted by Veselago, the reversal of the momentum density of light is still experimentally unverified. The reversal of the momentum density of light to be opposite to the electromagnetic (EM) energy flux density predicted by Veselago is counter-intuitive. It is well known that any field, particle, or quasiparticle, having positive energy propagating in one direction, \emph{must have} its momentum in the \emph{same direction}. Accordingly, the particle must fulfill the standard momentum per energy ratio $\mathbf{p}/E=\mathbf{v}_\mathrm{g}/c^2$ of relativistic mechanics \cite{Landau1989}, where $\mathbf{v}_\mathrm{g}$ is the group velocity and $c$ is the speed of light in vacuum. Therefore, if the total momentum of the coupled field-material state of light in NIMs is truly negative, the light in NIMs should be described by \emph{negative-energy quasiparticles}, illustrated in Fig.~\ref{fig:illustration}. The energy of these quasiparticles would not be equal to the known photon energy $\hbar\omega_0$, which is positive. This raises a question how these quasiparticles emerge when a photon is transferred from vacuum to a NIM in such a way that both the photon energy and momentum are transferred to the quasiparticle and the interface layer of the NIM crystal. The present work answers these questions by providing a detailed position- and time-dependent theory and the related computer simulations of the energy and momentum of light in NIMs.

In recent literature on optical forces in inhomogeneous materials like NIMs, it is typical to assume the Maxwell or Helmholtz stress tensor and further take the \emph{time average} of the related optical force density \cite{Sun2015,Wang2016a,Wang2016b,Rakich2010,Rakich2011,Pernice2009}. Homogenization of material parameters and fields may also be applied \cite{Sun2015,Wang2016a,Costa2011,Costa2009,Silverinha2007}. On the experimental side, the optical forces on metasurfaces have been very recently applied to drive microvehicles \cite{Andren2020}, but their use has not yet reached phenomena of light propagating inside NIM structures. The non-averaged \emph{exact position- and time-dependent} optical force and momentum densities of light in NIMs can be obtained by solving simultaneously Maxwell's equations for the field and the Newtonian dynamics of the solid, neutral material. The full picture of the energy and momentum fluxes associated with optical fields in NIMs can be obtained after solving the coupled dynamical equations.

In this work, we use the covariant mass-polariton (MP) theory of light \cite{Partanen2021b,Partanen2022a,Partanen2017c,Partanen2017e,Partanen2019a,Partanen2019b} to analyze the energy and momentum of light in NIMs. We also discuss experimental discovery of reduced, $p<p_0$, or even negative, $p<0$, radiation pressure, where $p$ and $p_0$ denote the components of total momentum of light in the NIM and in vacuum in the direction of propagation. The MP theory describes fields by using conventional Maxwell's equations, which explain the experimentally verified negative refraction and the inverse Doppler effect. However, our approach is fundamentally different from previous theories of negative radiation pressure in NIMs since it describes light in NIMs as coupled field-material MP states of Fig.~\ref{fig:illustration}. In the MP states of NIMs, the \emph{positive} EM energy and momentum densities are accompanied by \emph{positive or negative} field-driven energy and momentum densities of the material constituting an atomic mass density wave (MDW). We also show that, in the quantum picture, light in NIMs must be described as MP quasiparticles, whose total energy and momentum can become negative. The theory applies to both hypothetical homogeneous and conventional inhomogeneous NIMs structures. It shows that the total momentum of light in NIMs is typically smaller than the momentum of light in vacuum, and depending on the structure, it can even become negative. This contrasts to normal materials, for which the total momentum is always greater than its vacuum value. Our results demonstrate that the MP theory provides a versatile tool for engineering of optical forces in arbitrary NIMs structures and for designing of measurements, where the reduced or negative radiation pressure can be experimentally verified.

\section{Results}

\noindent\textbf{Momentum and energy densities of the EM field.}
We approach the description of light in arbitrary lossless photonic crystals from the classical field theory point of view assuming that the material is at rest in the absence of light. The energy flux density described by the Poynting vector $\mathbf{S}=\mathbf{E}\times\mathbf{H}$, where $\mathbf{E}$ and $\mathbf{H}$ are the electric and magnetic fields, is known to be continuous over material interfaces. This leads to a continuous EM contribution to the momentum density of light, given by \cite{Landau1984,Jackson1999}
\begin{equation}
 \mathbf{G}_\mathrm{EM}=\frac{\mathbf{E}\times\mathbf{H}}{c^2}.
 \label{eq:GEM}
\end{equation}

As discussed in Methods, the energy density of the EM field in the MP theory of light in dispersive media, solved from the conservation law of energy, is given by \cite{Partanen2022a}
\begin{equation}
 W_\mathrm{EM}=\frac{1}{2}\Big(\varepsilon\mathbf{E}^2+\mu\mathbf{H}^2
 +\omega_0\frac{d\varepsilon}{d\omega_0}\langle\mathbf{E}^2\rangle+\omega_0\frac{d\mu}{d\omega_0}\langle\mathbf{H}^2\rangle\Big).
 \label{eq:WEM}
\end{equation}
Here $\omega_0$ is the central angular frequency of the \emph{narrow frequency band}, and $\varepsilon=\varepsilon(\omega_0)$ and $\mu=\mu(\omega_0)$ are the permittivity and permeability of the material at $\omega_0$. For a narrow frequency band, the permittivity can be approximated by the first two terms of its Taylor series $\varepsilon(\omega)\approx\varepsilon(\omega_0)+\frac{\partial\varepsilon(\omega_0)}{\partial\omega_0}(\omega-\omega_0)$. A similar expansion is valid for the permeability. The angle brackets in Eq.~\eqref{eq:WEM} are defined by $\langle\mathbf{E}^2\rangle=\frac{1}{2}[\mathbf{E}^2+(\frac{1}{\omega_0}\frac{\partial}{\partial t}\mathbf{E})^2]$ and $\langle\mathbf{H}^2\rangle=\frac{1}{2}[\mathbf{H}^2+(\frac{1}{\omega_0}\frac{\partial}{\partial t}\mathbf{H})^2]$. The brackets here and below are also equal to the time average of the pertinent quantities over the harmonic cycle, e.g., $\langle\mathbf{E}^2(\mathbf{r},t)\rangle=\frac{1}{T}\int_{-T/2}^{T/2}\mathbf{E}^2(\mathbf{r},t+t')dt'$, where $T=2\pi/\omega_0$ is the length of the harmonic cycle.

\vspace{0.5cm}
\noindent\textbf{Optical force density.}
The electric and magnetic fields $\mathbf{E}$ and $\mathbf{H}$ and corresponding flux densities $\mathbf{D}$ and $\mathbf{B}$ of light in NIMs can be solved from Maxwell's equations. What is less well known is the dynamics of atoms in NIMs under the influence of the instantaneous optical force density. The microscopic form of the optical force density in the MP theory of light accounting for the gradients of material parameters in inhomogeneous dispersive materials is given by \cite{Partanen2022a}
\begin{align}
 \mathbf{f}_\mathrm{opt} &=-\frac{\partial\mathbf{G}_\mathrm{EM}}{\partial t}
 -\nabla\cdot\boldsymbol{\mathcal{T}}_\mathrm{EM}.
 \label{eq:fopt}
\end{align}
In Eq.~\eqref{eq:fopt}, $\boldsymbol{\mathcal{T}}_\mathrm{EM}$ is the EM stress tensor which, in the MP theory of ligh, is given by \cite{Partanen2022a}
\begin{equation}
 \boldsymbol{\mathcal{T}}_\mathrm{EM}=\frac{1}{2}\Big[(\mathbf{E}\cdot\mathbf{D}+\mathbf{H}\cdot\mathbf{B})\mathbf{I}-\mathbf{E}\otimes\mathbf{D}-\mathbf{D}\otimes\mathbf{E}-\mathbf{H}\otimes\mathbf{B}-\mathbf{B}\otimes\mathbf{H}\Big].
 \label{eq:TEM}
\end{equation}
Here $\mathbf{I}$ is the $3\times3$ unit matrix and $\otimes$ denotes the outer product of vectors.

\vspace{0.5cm}
\noindent\textbf{Momentum and energy densities of the atomic MDW and the coupled MP state of light.}
When light is interacting with a lossless material, the dynamics of the material is described solely by the Newton's equation. Assuming that there are no other forces except the optical force, including both interface and bulk forces, Newton's equation of motion for the material is given by \cite{Penfield1967,Partanen2021b}
\begin{equation}
 n_\mathrm{a}\frac{d\mathbf{p}_\mathrm{a}}{dt}
 =\mathbf{f}_\mathrm{opt}.
 \label{eq:Newton}
\end{equation}
Here $\mathbf{p}_\mathrm{a}$ is the momentum of a single atom and $n_\mathrm{a}$ is the number density of atoms. Since the atomic velocities are nonrelativistic, the atomic momentum in Eq.~\eqref{eq:Newton} can be written classically $\mathbf{p}_\mathrm{a}= m_0\mathbf{v}_\mathrm{a}$, where $m_0$ is the mass and $\mathbf{v}_\mathrm{a}$ is the velocity of the atom. The momentum density of the MDW, driven forward by light, is then given by \cite{Partanen2021b}
\begin{equation}
\mathbf{G}_\mathrm{MDW}=m_0\mathbf{v}_\mathrm{a}n_\mathrm{a}=\int_{-\infty}^t\mathbf{f}_\mathrm{opt}dt'.
 \label{eq:GMDW}
\end{equation}
The energy density of the atomic MDW in the approximation of negligible kinetic energy of atoms is given by
\begin{equation}
 W_\mathrm{MDW}=m_0c^2(n_\mathrm{a}-n_\mathrm{a0}),
 \label{eq:WMDW}
\end{equation}
where $n_\mathrm{a0}$ is the atomic number density of the material before the optical force has started to displace the atoms.

In locally homogeneous regions of space, the momentum and energy densities of the atomic MDW, resulting from Eqs.~\eqref{eq:GMDW} and \eqref{eq:WMDW} and the optical force density in Eq.~\eqref{eq:fopt}, are given by \cite{Partanen2022a}
\begin{equation}
 \mathbf{G}_\mathrm{MDW}=(n_\mathrm{p}^2-1)\mathbf{G}_\mathrm{EM}+n_\mathrm{p}(n_\mathrm{g}-n_\mathrm{p})\langle\mathbf{G}_\mathrm{EM}\rangle,
 \label{eq:GMDWh}
\end{equation}
\begin{equation}
 W_\mathrm{MDW}=(n_\mathrm{p}^2-1)W_\mathrm{EM}+n_\mathrm{p}(n_\mathrm{g}-n_\mathrm{p})\langle W_\mathrm{EM}\rangle.
 \label{eq:WMDWh}
\end{equation}
Here $n_\mathrm{p}$ and $n_\mathrm{g}$ are the local phase and group refractive indices of the material at the central angular frequency of the narrow frequency band. In Eq.~\eqref{eq:GMDWh}, the angle bracket is given by $\langle\mathbf{G}_\mathrm{EM}\rangle=\frac{1}{2}[\mathbf{E}\times\mathbf{H}+(\frac{1}{\omega_0}\frac{\partial}{\partial t}\mathbf{E})\times(\frac{1}{\omega_0}\frac{\partial}{\partial t}\mathbf{H})]/c^2$. This time- and position-dependent quantity is also equal to the time average of the EM momentum density over the harmonic cycle.

The total momentum density of the MP state of light is a sum of the EM momentum density in Eq.~\eqref{eq:GEM} and the momentum density of the MDW in Eq.~\eqref{eq:GMDW} as
\begin{equation}
 \mathbf{G}_\mathrm{MP}=\mathbf{G}_\mathrm{EM}+\mathbf{G}_\mathrm{MDW}.
 \label{eq:GMP1}
\end{equation}
Correspondingly, the total energy density of the coupled MP state of light is a sum of the EM energy density in Eq.~\eqref{eq:WEM} and the energy density of the MDW in Eq.~\eqref{eq:WMDW} as
\begin{equation}
 W_\mathrm{MP}=W_\mathrm{EM}+W_\mathrm{MDW}.
 \label{eq:WMP1}
\end{equation}
See Methods for a brief comparison of the total momentum density of light in the MP theory in Eq.~\eqref{eq:GMP1} with the total momentum density of light used by Veselago \cite{Veselago1968}.

\vspace{0.5cm}
\noindent\textbf{Light in a hypothetical homogeneous NIM.}
Here we apply the theory to homogeneous NIMs, even though such materials are not presently available. The total momentum and energy of the atomic MDW are volume integrals of the classical momentum and energy densities of atoms in Eqs.~\eqref{eq:GMDW} and \eqref{eq:WMDW}. For a homogeneous medium, one thus obtains \cite{Partanen2021b}
\begin{equation}
 \mathbf{p}_\mathrm{MDW}=\int\mathbf{G}_\mathrm{MDW}d^3r=(n_\mathrm{p}n_\mathrm{g}-1)\mathbf{p}_\mathrm{EM},
 \label{eq:pMDW}
\end{equation}
\begin{equation}
 E_\mathrm{MDW}=\int W_\mathrm{MDW}d^3r=(n_\mathrm{p}n_\mathrm{g}-1)E_\mathrm{EM}.
 \label{eq:EMDW}
\end{equation}
Here $\mathbf{p}_\mathrm{EM}=\int\mathbf{G}_\mathrm{EM}d^3r$ and $E_\mathrm{EM}=\int W_\mathrm{EM}d^3r$ are the total EM momentum and energy of the light pulse. The last forms of Eqs.~\eqref{eq:pMDW} and \eqref{eq:EMDW} are straightforwardly obtained from Eqs.~\eqref{eq:GMDWh} and \eqref{eq:WMDWh} or from the MP quasiparticle model for dispersive media \cite{Partanen2017e}.

\begin{figure}
\centering
\includegraphics[width=0.80\columnwidth]{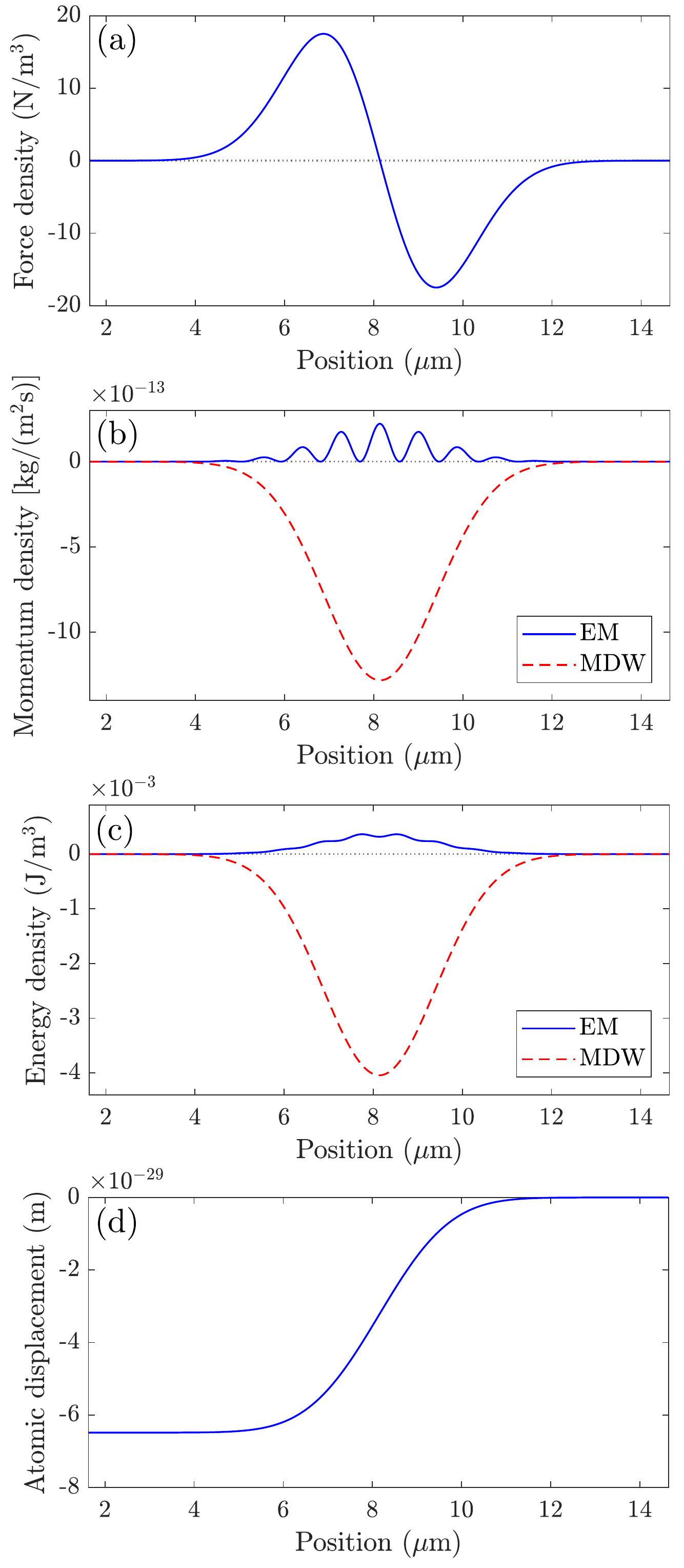}
\caption{\label{fig:simulationhomog}
\textbf{Simulation of light in a homogeneous NIM.} The total energy and momentum of light are shared between the EM field and the material atoms. (a) Optical force density, (b) EM and MDW momentum densities, (c) EM and MDW energy densities, and (d) the atomic displacement for a one-dimensional Gaussian light pulse propagating to the right at a fixed instance of time. The atomic displacement is here very small, but it can be \emph{nanometers} for pulses that are longer and have higher intensity \cite{Partanen2017e}.}
\end{figure}

By summing the EM field and atomic MDW contributions, one obtains the total momentum and energy of the coupled MP state of light in homogeneous NIMs as
\begin{equation}
 \mathbf{p}_\mathrm{MP}=\mathbf{p}_\mathrm{EM}+\mathbf{p}_\mathrm{MDW}=n_\mathrm{p}n_\mathrm{g}\mathbf{p}_\mathrm{EM},
 \label{eq:pMP}
\end{equation}
\begin{equation}
 E_\mathrm{MP}=E_\mathrm{EM}+E_\mathrm{MDW}=n_\mathrm{p}n_\mathrm{g}E_\mathrm{EM}.
 \label{eq:EMP}
\end{equation}
The total MP momentum in Eq.~\eqref{eq:pMP} is \emph{directed opposite} to the volume integral of the Poynting vector and the related total EM momentum for NIMs with $n_\mathrm{p}<0$, and it is discussed further in Methods. One can conclude that the optical force density in Eq.~\eqref{eq:fopt} on average reduces the number density of atoms from $n_\mathrm{a0}$ to $n_\mathrm{a}$, and thus, makes the total energy of the coupled field-material state of light in Eq.~\eqref{eq:EMP} \emph{negative}. Therefore, in the MP theory of light, the total energy and the component of the momentum in the direction of propagation of light are both negative in homogeneous NIMs.

We visualize the dynamics of the material and the emergence of the atomic MDW consisting of local rarefications of the material density by using a Gaussian light pulse and the simulation parameters of a hypothetical homogeneous NIM described in Methods. The Gaussian light pulse has a central vacuum wavelength of $\lambda_0=1746.7$ nm, the peak electric field amplitude of $E_0=2744.9$ V/m, and the relative spectral width of $\Delta\lambda_0/\lambda_0=0.015$. Figure \ref{fig:simulationhomog}(a) shows the optical force density of Eq.~\eqref{eq:fopt} for the simulated Gaussian light pulse at an instance of time when the center of the light pulse is at position $x=8$ $\mu$m. Figure \ref{fig:simulationhomog}(b) presents the EM and atomic MDW momentum densities of Eqs.~\eqref{eq:GEM} and \eqref{eq:GMDW}. Figure \ref{fig:simulationhomog}(c) depicts the EM and atomic MDW energy densities of Eqs.~\eqref{eq:WEM} and \eqref{eq:WMDW}. The values of the EM quantities are \emph{positive} at all positions and times while the MDW quantities are \emph{negative}. The atoms are \emph{drawn backward} and thus, the density of the material is locally rarefied. The atomic displacement $r_\mathrm{a}=\int_{-\infty}^tv_\mathrm{a}dt'$ is depicted in Fig.~\ref{fig:simulationhomog}(d). It shows how the light pulse leaves atoms behind it displaced backward. The atomic displacement is very small, but it can be \emph{nanometers} for pulses that are longer and have higher intensity as shown in Fig.~7 of Ref.~\cite{Partanen2017e}.

\vspace{0.5cm}
\noindent\textbf{Consistency with the principles of relativistic mechanics.}
We next show by a quasiparticle analysis that the positive total energy density of light of earlier works \cite{Veselago1968,Ward2005,Barnett2015} on homogeneous NIMs is very challenging since, in relativistic mechanics, the momentum per energy ratio must be given by $\mathbf{p}/E=\mathbf{v}_\mathrm{g}/c^2$, where $\mathbf{v}_\mathrm{g}$ is the group velocity \cite{Landau1989}. Using Eqs.~\eqref{eq:pMP} and \eqref{eq:EMP} and the EM energy $E_\mathrm{EM}=\int W_\mathrm{EM}d^3r=\hbar\omega_0$ normalized for a single-photon, the MP energy and momentum in a homogeneous NIM are given by $E_\mathrm{MP}=\int W_\mathrm{MP}d^3r=n_\mathrm{p}n_\mathrm{g}\hbar\omega_0$ and $\mathbf{p}_\mathrm{MP}=\int\mathbf{G}_\mathrm{MP}d^3r=(n_\mathrm{p}\hbar\omega_0/c)\mathbf{v}_\mathrm{g}/|\mathbf{v}_\mathrm{g}|$ \cite{Partanen2017e,Partanen2021b}. Thus, the MP quasiparticle fulfills the standard momentum per energy ratio $\mathbf{p}/E=\mathbf{v}_\mathrm{g}/c^2$ of relativistic mechanics \cite{Landau1989}. This ratio is violated in conventional theories of light in homogeneous NIMs \cite{Veselago1968,Ward2005,Barnett2015} pinpointing the importance of the MDW energy included in the present theory.

\begin{figure}
\centering
\includegraphics[width=\columnwidth]{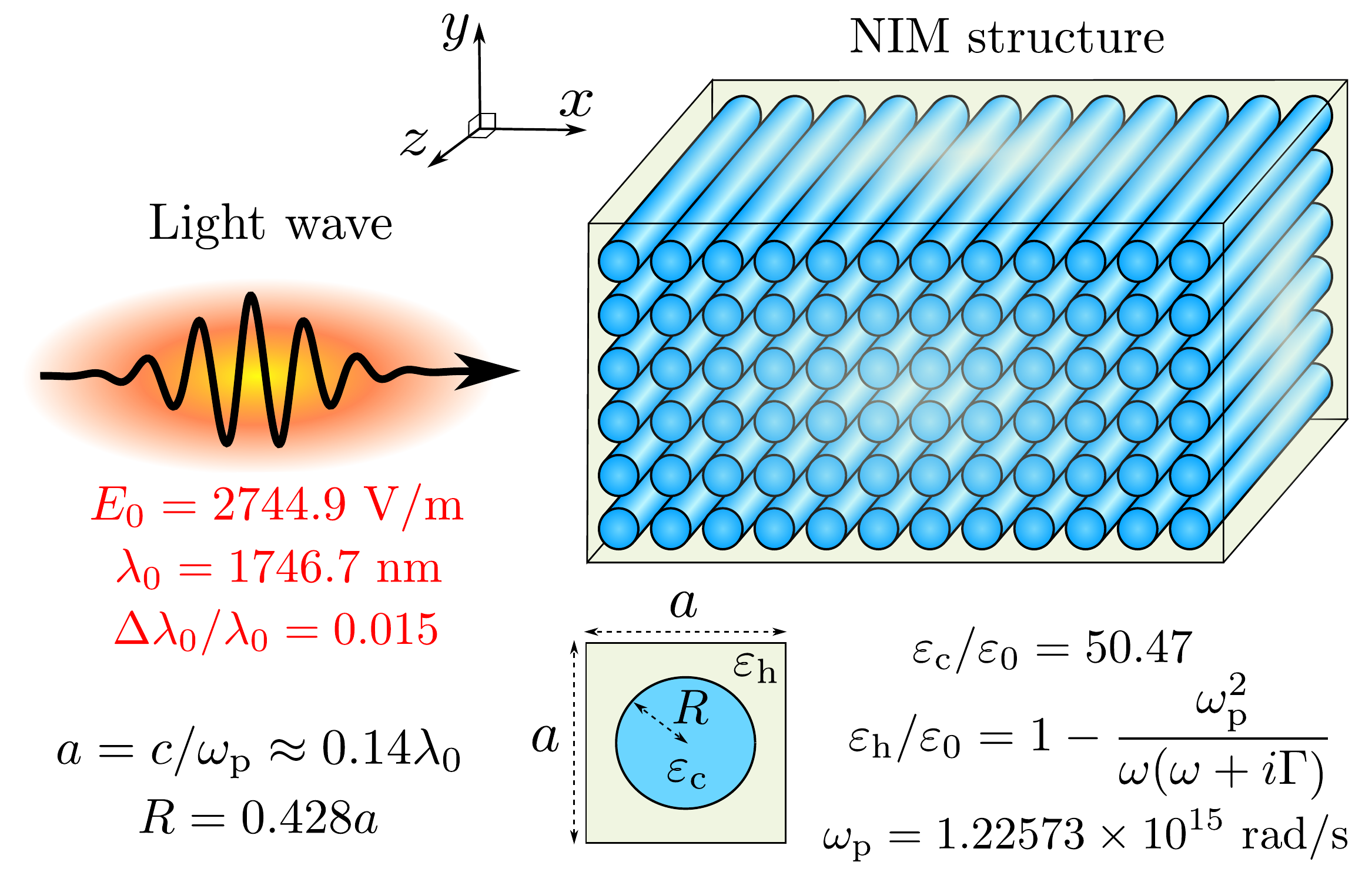}
\caption{\label{fig:inhomogeneoussetup}
\textbf{Schematic illustration of light incident to an inhomogeneous NIM block.} The NIM structure consists of dielectric cylinders embedded in a plasmonic host material. Light is incident from the left and it propagates through the structure with minimal backward reflections.}
\end{figure}

\vspace{0.5cm}
\noindent\textbf{Light in an inhomogeneous NIM.}
In inhomogeneous NIMs, i.e., in all NIM structures that have been experimentally realized so far, the total momentum and energy of light include momentum and energy terms related to the gradients of material parameters and internal material interfaces of the NIM structure. This means that the MDW in typical NIM geometries is split into the volume and internal interface contributions. Thus, for the MP state of light, we obtain
\begin{equation}
 \mathbf{p}_\mathrm{MP}=\mathbf{p}_\mathrm{EM}+\mathbf{p}_\mathrm{MDW}^\mathrm{(vol)}+\mathbf{p}_\mathrm{MDW}^\mathrm{(int)},
 \label{eq:pMPih}
\end{equation}
\begin{equation}
 E_\mathrm{MP}=E_\mathrm{EM}+E_\mathrm{MDW}^\mathrm{(vol)}+E_\mathrm{MDW}^\mathrm{(int)}.
 \label{eq:EMPih}
\end{equation}
These relations are more complex than the corresponding relations for a homogeneous NIM in Eqs.~\eqref{eq:pMP} and \eqref{eq:EMP}. Consequently, the values of $\mathbf{p}_\mathrm{MP}$ and $E_\mathrm{MP}$ following from Eqs.~\eqref{eq:pMPih} and \eqref{eq:EMPih} for general NIM structures can be either positive or negative. Whether it is possible to construct effective parameters in terms of which the total momentum and energy of light in an inhomogeneous NIM can be written using the EM quantities, as in Eqs.~\eqref{eq:pMP} and \eqref{eq:EMP} for a homogeneous NIM, is an open question left as a topic of further work. In any case, it can be shown that the MP energy and momentum in Eqs.~\eqref{eq:pMPih} and \eqref{eq:EMPih} fulfill the standard momentum per energy ratio $\mathbf{p}/E=\mathbf{v}_\mathrm{g}/c^2$ of relativistic mechanics \cite{Landau1989}.

The internal interface MDW momentum and energy contributions $\mathbf{p}_\mathrm{MDW}^\mathrm{(int)}$ and $E_\mathrm{MDW}^\mathrm{(int)}$ in Eqs.~\eqref{eq:pMPih} and \eqref{eq:EMPih} do not include contributions from the external interfaces of the NIM structure. The excluded external interface contributions $\mathbf{p}_\mathrm{interface}$ and $E_\mathrm{interface}$ are left at the interface when the light pulse enters the NIM structure. They decay afterward by internal phonon relaxation of the structure. These terms also account for the partial reflection taking place at the interface. The terms $\mathbf{p}_\mathrm{interface}$ and $E_\mathrm{interface}$ are essential in the conservation laws when light enters or exits the NIM structure. We also note that, in the present work, the conventional boundary conditions of the EM fields are used at all material interfaces. This assumes that the material boundaries are sharp. In general, this is not the case, and the surfaces can be rough, for example. If the structure size is made close to the atomic scale, the surfaces cannot be sharp, and it is necessary to account for the related surface effects.

An example of an inhomogeneous NIM structure is schematically illustrated in Fig.~\ref{fig:inhomogeneoussetup}. The structure consists of dielectric cylinders embedded in a plasmonic host material. The propagation of light in this kind of a structure has been previously theoretically investigated, e.g., by Costa \emph{et al.} \cite{Costa2011}. In our setting, there are approximately seven unit cells in a single vacuum wavelength of the light pulse. More detailed material and structure parameters are shown in Fig.~\ref{fig:inhomogeneoussetup} and discussed in Methods. The parameters of the incident Gaussian light pulse are the same as those in the case of a homogeneous NIM above.

\begin{figure}
\centering
\includegraphics[width=\columnwidth]{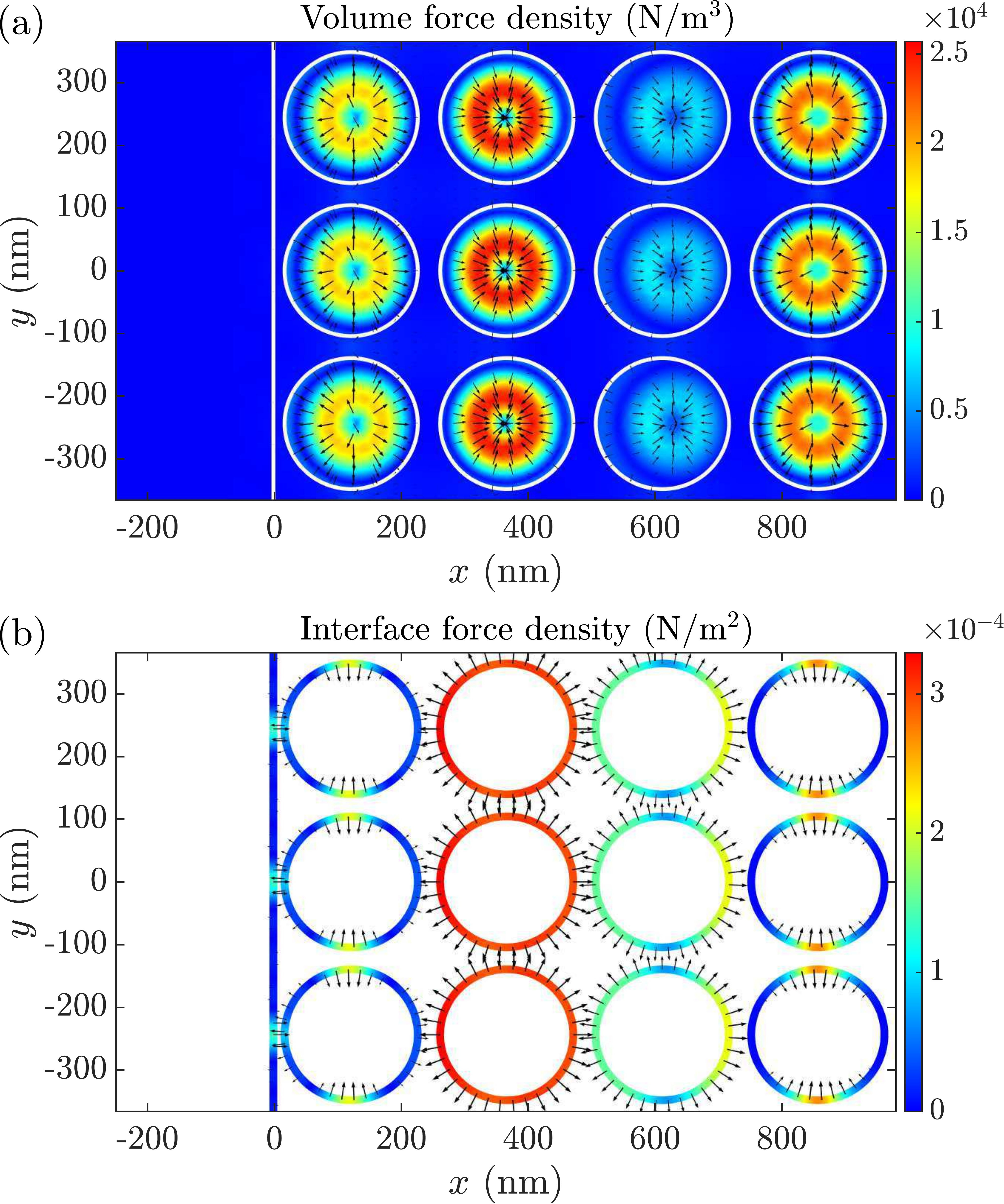}
\caption{\label{fig:forcedensities}
\textbf{Volume and interface force densities in an inhomogeneous NIM.} The instantaneous (a) volume and (b) interface force densities as a function of position around the entry interface of the NIM structure. The snapshot is taken at the instance of time in which the energy density maximum of the Gaussian light pulse incident from the left is at $x=0$. The linearly scaled color indicates the magnitude, and the logarithmically scaled arrows indicate the direction of the force densities. White solid lines depict material interfaces.}
\end{figure}

\begin{table}
 \setlength{\tabcolsep}{3.0pt}
 \renewcommand{\arraystretch}{2.4}
 \caption{\label{tbl:table}
 \textbf{Momentum components after the light pulse has entered the NIM.} When the light pulse enters the NIM, the incident EM momentum component $p_0$ in the direction of propagation is split to partial momentum components carried by the reflected EM field ($p_\mathrm{EM}^\mathrm{(refl)}$), EM field in NIM ($p_\mathrm{EM}^\mathrm{(NIM)}$), atomic MDW volume contribution ($p_\mathrm{MDW}^\mathrm{(vol)}$), MDW contribution of internal interfaces ($p_\mathrm{MDW}^\mathrm{(int)}$), and the atoms of the vacuum-NIM interface ($p_\mathrm{interface}$). The table shows the momentum contributions for homogeneous and inhomogeneous NIMs studied. These two cases have the same effective phase and group refractive indices $n_\mathrm{p}=-1$ and $n_\mathrm{g}=10.52$. Within the relative numerical accuracy of 0.2\% of our simulations, the sum of the partial momenta per $p_0$ is equal to unity.}
\begin{tabular}{cccccc}
   \hline\hline
   & $\dfrac{p_\mathrm{EM}^\mathrm{(refl)}}{p_0}$ & $\dfrac{p_\mathrm{EM}^\mathrm{(NIM)}}{p_0}$ & $\dfrac{p_\mathrm{MDW}^\mathrm{(vol)}}{p_0}$ & $\dfrac{p_\mathrm{MDW}^\mathrm{(int)}}{p_0}$ & $\dfrac{p_\mathrm{interface}}{p_0}$ \\[4pt]
   \hline
   $Homog.$ & $0$ & $0.0951$ & $-1.0951$ & $0$ & $2$ \\[-21pt]
   & & \multicolumn{3}{c}{$\underbrace{\hspace{3.5cm}}_{\displaystyle p_\mathrm{MP}=-p_0}$} & \\[14pt]
   $Inhomog.$ & $-0.0007$ & $0.0954$ & $0.6743$ & $-0.4527$ & $0.6820$ \\[-21pt]
   & & \multicolumn{3}{c}{$\underbrace{\hspace{3.5cm}}_{\displaystyle p_\mathrm{MP}=0.317p_0}$} & \\[14pt]
   \hline\hline
 \end{tabular}
\end{table}

\begin{figure*}
\centering
\includegraphics[width=\textwidth]{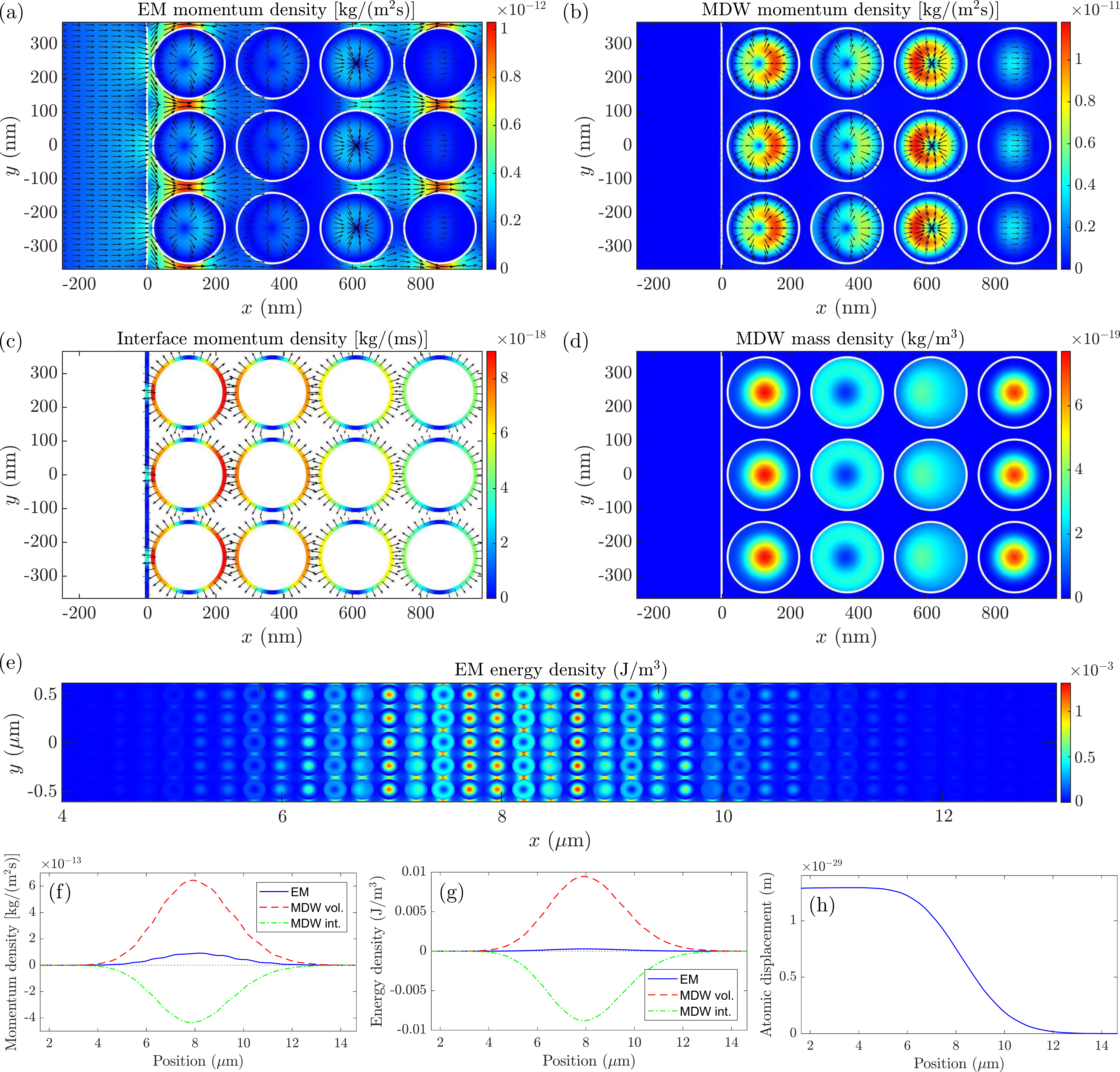}
\caption{\label{fig:simulationinhomog}
\textbf{Simulation of light in an inhomogeneous NIM.} The instantaneous (a) EM volume momentum density, (b) the atomic MDW volume momentum density, (c) the surface momentum density of interface atoms, and (d) the mass density of the atomic MDW as a function of position around the entry interface of the NIM structure. The snapshot is taken at the instance of time when the energy density maximum of the Gaussian light pulse incident from the left is at $x=0$. The linearly scaled color indicates the magnitude, and the logarithmically scaled arrows indicate the direction of the momentum densities. White solid lines depict material interfaces. (e) The EM energy density, (f) spatially averaged EM, MDW volume, and MDW internal interface momentum densities, (g) spatially averaged EM, MDW volume, and MDW internal interface energy densities, and (h) spatially averaged atomic displacement at an instance of time after the pulse has penetrated inside the NIM.}
\end{figure*}

Figure \ref{fig:forcedensities} presents the optical volume and interface force densities resulting from the Gaussian light pulse. The snapshot is taken at the instance of time when the center of the Gaussian pulse is at the position of the entry interface of the NIM structure. Figures \ref{fig:simulationinhomog}(a)--(d) show the position dependencies of the EM momentum density, MDW volume momentum density, interface momentum density, and the MDW mass density, respectively, at the instance of time corresponding to the force densities in Fig.~\ref{fig:forcedensities}. The EM momentum density in Fig.~\ref{fig:simulationinhomog}(a) consists of the incident and transmitted contributions. The reflection is close to zero for the NIM structure studied. The atomic MDW momentum density in Fig.~\ref{fig:simulationinhomog}(b) is zero on the left since there is no MDW in vacuum. It has the highest values in the dielectric cylinders, but it is also locally nonzero in the plasmonic material, where it is, however, so much smaller than in dielectric cylinders that this cannot be seen in the color scale of the figure. The interface momentum density in Fig.~\ref{fig:simulationinhomog}(c) is directed close to the normal of the interfaces. Part of this momentum density stays in the vicinity of the vacuum-NIM interface when the light pulse has passed while part of it propagates with light inside the NIM. Figure \ref{fig:simulationinhomog}(d) presents the mass density of the atomic MDW, which is driven forward by the optical volume force density. Like the MDW momentum density in Fig.~\ref{fig:simulationinhomog}(b), the MDW mass density is also locally small but nonzero in the plasmonic material. The MDW mass density follows the Gaussian envelope of the light pulse. Apart from different relative magnitudes in the dielectric cylinders and in the plasmonic host material, it reminds the EM energy density.

\clearpage

\begin{figure*}
\centering
\includegraphics[width=0.99\textwidth]{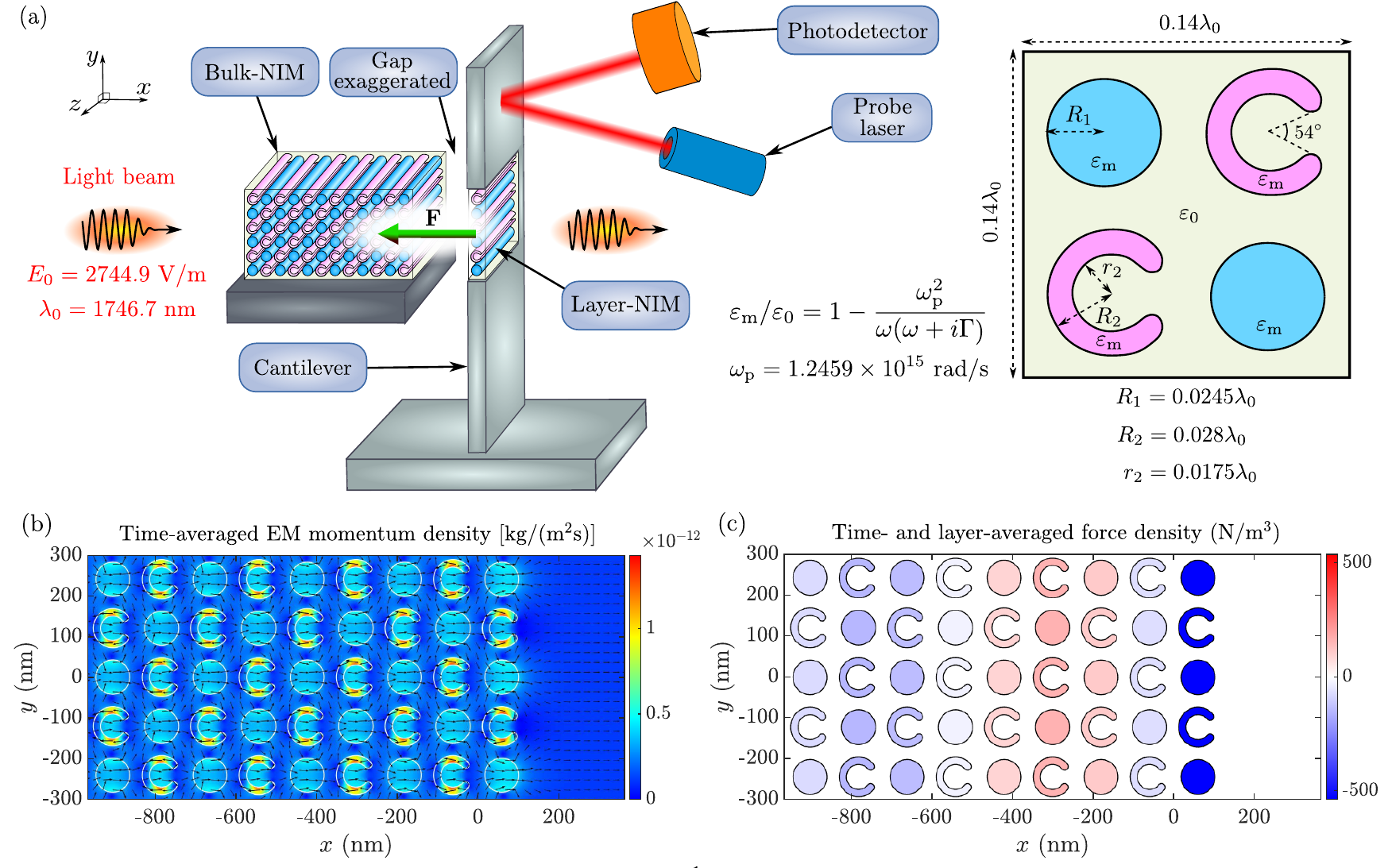}
\caption{\label{fig:setup}
\textbf{Schematic experimental setup and simulation of reduced or negative radiation pressure.} (a) Schematic setup for the measurement of the force on the interface where the light beam exits the NIM crystal. The NIM crystal is fabricated on a substrate so that the cylinders and c-shaped rods are fixed from their ends (not shown). The NIM crystal is then split into bulk-NIM and layer-NIM parts. The bulk-NIM is attached to a rigid supporting structure while the layer-NIM is attached to a cantilever. The gap between the bulk- and layer-NIMs is exaggerated to make it visible. In the proposed experiment, this gap is equal to the gaps inside layers of the bulk-NIM. The cantilever is able to bend as a result of the optical force. The nanoscale bending ($\ll a$) of the cantilever  is recorded by monitoring the reflection of a probe laser beam using a photodetector. The right inset shows the NIM structure used in our example. (b) Time-averaged EM momentum density of a continuous wave light beam propagating through the NIM.  The linearly scaled color indicates the magnitude, and the logarithmically scaled arrows indicate the direction. (c) The $x$ component of the total optical force density averaged over the harmonic cycle and over the vertical layers of the NIM. The bulk-NIM is at $x<0$ and the layer-NIM is at $x>0$.}
\end{figure*}

The EM energy density is depicted in a larger scale in Fig.~\ref{fig:simulationinhomog}(e) at a later instance of time when the trailing edge of the pulse has also entered the NIM. At this instance of time, we have also plotted the spatially averaged EM, MDW volume, and internal interface momentum and energy densities, and the atomic displacement, in Figs.~\ref{fig:simulationinhomog}(f)--(h), respectively. The spatial averaging is made over a distance of one wavelength. These figures allow for transparent comparison with the case of a homogeneous NIM in Fig.~\ref{fig:simulationhomog}. It is seen that there are negative contributions in the momentum and energy densities, which originate from the internal interfaces in the MDW. However, in the present case, the volume contributions of the MDW are more positive and the net effect is a slight densification of the material, which contrasts to the rarefication seen in Fig.~\ref{fig:simulationhomog}. From the negative internal interface contribution, it however, results that the net momentum propagating with the light pulse in the inhomogeneous NIM is smaller than the momentum of the pulse in vacuum. This is discussed in more detail for the integrals of the momentum density contributions below. Apart from the slight spreading of the pulse in the nonlinear dispersion of the inhomogeneous NIM, the spatially averaged EM momentum and energy densities in Figs.~\ref{fig:simulationinhomog}(f) and (g) are approximately equal to what would be obtained by correspondingly averaging the EM quantities of the homogeneous case in Figs.~\ref{fig:simulationhomog}(b) and (c).

When the light pulse enters the NIM, the incident EM momentum component $p_0$ is split to partial momenta carried by the reflected and transmitted parts of the EM field, the atomic MDW, and the interface atoms. The comparison of these momenta between the homogeneous and inhomogeneous NIM cases studied above is presented in Table \ref{tbl:table}. The total MP momentum component of the coupled field-material state of light in both cases is smaller than $p_0$. In the homogeneous case, it is negative while, in the inhomogeneous case, it is positive. Since the total MP momentum component is smaller than $p_0$, the interface momentum is directed toward the NIM due to the conservation law of momentum. The same applies at the interface where the light pulse exits the NIM.

\vspace{0.5cm}
\noindent\textbf{Experimental detection of reduced or negative radiation pressure.}
A schematic experimental setup for probing radiation pressure of light in NIMs is illustrated in Fig.~\ref{fig:setup}(a). The setup enables the measurement of the optical force produced when light exits the NIM structure. The atomic MDW densification or rarefication waves are also present in this setup which, however, focuses on the force measurement \cite{Melcher2014,Evans2014,Kleckner2006,Ma2015,Ma2018,Partanen2021a,Partanen2020b,Vaskuri2021}. In contrast to the NIM in Fig.~\ref{fig:inhomogeneoussetup}, here we use a NIM where the host material is vacuum and there are cylinders and c-shaped rods made of a plasmonic material. See Refs.~\cite{Kadic2019,Shalaev2007} for examples of artificial structures that can be fabricated with presently available technologies. The NIM is split into bulk-NIM and layer-NIM parts. The layer-NIM is attached to a cantilever, which is able to bend. Therefore, the force experienced by the layer-NIM can be measured by detecting the bending of the cantilever under the continuous-wave illumination of the NIM. This bending can be detected with a probe laser beam and a photodetector utilizing the optics of an atomic force microscope. Cantilever displacements of fractions of a nanometer are measurable. The setup of Fig.~\ref{fig:setup}(a) is not optimized for the actual measurement but it is meant to provide an idea of the measurement that can be realized with various kinds of NIM structures.

Figure \ref{fig:setup}(b) presents the time-averaged EM momentum density. It shows that, on average, the energy and momentum fluxes of the EM field are directed forward in the cylinders and, counterintuitively, backward in the c-shaped rods. Figure \ref{fig:setup}(c) depicts the force density averaged both over the harmonic cycle and over vertical layers of the NIM structure. The layer-NIM attached to the cantilever experiences the strongest force and it is directed backward, i.e., the radiation pressure is reduced or negative. Qualitatively, the same result applies to various NIM structures in which the total momentum of light is smaller than that of vacuum, i.e., the momentum of light does not need to be negative. The force directed inward the NIM contrasts the force in normal materials where it is directed outward \cite{Astrath2014}. Since the host material of the NIM used in the setup is vacuum, the conventional Maxwell stress tensor of vacuum could also be applied giving results equal to those of us.

\vspace{-0.4cm}
\section{Conclusions}

In conclusion, in contrast to normal materials, the total momentum of light in NIMs can be smaller than the momentum of light in vacuum. The radiation pressure in NIMs can even become negative depending on the subwavelength structure of the material. In NIMs exhibiting negative radiation pressure, the negative total momentum of light results from the atomic mass density rarefication waves caused by the optical force density. In the corresponding single light quantum picture, the MP quasiparticles have a negative total energy, which allows the total momentum of light to be directed opposite to the group velocity. The backward atomic velocities and the resulting mass density rarefication are real, experimentally measurable, and necessary for the fulfillment of the constant center-of-energy velocity of an isolated system. The theory applies to both hypothetical homogeneous and conventional inhomogeneous NIMs structures. We have presented a schematic experimental setup, which allows the measurement of the reduced or negative radiation pressure effect of light in NIMs. The present results demonstrate the versatile use of the MP theory of light for modeling optomechanics of light in general photonic crystal structures.

\vspace{0.8cm}\noindent\textbf{METHODS}\vspace{0.4cm}\\
\small
\noindent\textbf{Energy densities.}
In a lossless material, the energy density of the EM field can be solved from the local conservation law of energy, $\frac{1}{c^2}\frac{\partial}{\partial t}W_\mathrm{EM}+\nabla\cdot\mathbf{G}_\mathrm{EM}=-\mathbf{f}_\mathrm{opt}\cdot\mathbf{v}_\mathrm{a}$, where the right-hand side can be approximated by zero due to its smallness, giving $W_\mathrm{EM} =-c^2\int_{-\infty}^t\nabla\cdot\mathbf{G}_\mathrm{EM}dt'$ \cite{Partanen2021b,Philbin2011}. This integral expression then leads to the result, given in Eq.~\eqref{eq:WEM} \cite{Partanen2022a}. Solving the energy density of the MDW from the conservation law of energy correspondingly gives Eq.~\eqref{eq:WMDW}, which for homogeneous regions of space is equal to Eq.~\eqref{eq:WMDWh} \cite{Partanen2022a}.

\vspace{0.2cm}\noindent\textbf{Total momentum density of light.}
The instantaneous momentum density of the MP state of light in Eq.~\eqref{eq:GMP1} is not equal to the total momentum density of light used by Veselago \cite{Veselago1968}, i.e., $n_\mathrm{p}^2\mathbf{E}\times\mathbf{H}/c^2+\frac{1}{2}\mathbf{k}(\frac{\partial\varepsilon}{\partial\omega_0}\mathbf{E}^2+\frac{\partial\mu}{\partial\omega_0}\mathbf{H}^2)$, where $\mathbf{k}$ is the wave vector of light in the material. For homogeneous NIMs, one can, however, show that the harmonic cycle time averages of these momentum densities are equal. Veselago did not consider the possibility that part of the total momentum of light could be carried by the material. Therefore, his approach deviates from the MP theory. For homogeneous NIMs with $n_\mathrm{p}<0$, the MP momentum in Eq.~\eqref{eq:pMP} is directed opposite to the EM energy and momentum flux. This fundamentally originates from the backward-directed optical force in Eq.~\eqref{eq:fopt} and the resulting backward-directed atomic velocities and the MDW momentum density. For normal dispersive materials with $n_\mathrm{p}>0$, the total momentum of the MP state of light in Eq.~\eqref{eq:pMP} explains the results of the high-precision measurements of radiation pressure by Jones \emph{et al.} \cite{Jones1978}.

\vspace{0.2cm}\noindent\textbf{MP quasiparticles.}
In the single light quantum picture, the energy and momentum of the MP quasiparticle can be shown to be equal to the volume integrals of the classical energy and momentum densities normalized for the known single-photon EM energy $E_\mathrm{EM}=\int W_\mathrm{EM}d^3r=\hbar\omega_0$. For homogeneous NIMs, this leads to $E_\mathrm{MP}=\int W_\mathrm{MP}d^3r=n_\mathrm{p}n_\mathrm{g}\hbar\omega_0$ and $\mathbf{p}_\mathrm{MP}=\int\mathbf{G}_\mathrm{MP}d^3r=(n_\mathrm{p}\hbar\omega_0/c)\mathbf{v}_\mathrm{g}/|\mathbf{v}_\mathrm{g}|$ \cite{Partanen2017e,Partanen2021b}. Using the relativistic energy-momentum relation, given by $(m_\mathrm{MP}c^2)^2=E_\mathrm{MP}^2-(p_\mathrm{MP}c)^2$, one then obtains the MP quasiparticle rest mass as
$m_\mathrm{MP}=n_\mathrm{p}\sqrt{n_\mathrm{g}^2-1}\,\hbar\omega_0/c^2$. Therefore, for homogeneous NIMs, $m_\mathrm{MP}$ is negative. This result can also be obtained by using the relativistic MP quasiparticle model presented in Ref.~\cite{Partanen2017e}. The energy and momentum of the MP quasiparticle in all inertial frames are given by the same relativistic relations as for all other particles and quasiparticles with a nonzero rest mass. Thus, for the MP state of light, we have $E_\mathrm{MP}=\gamma_{\mathbf{v}_\mathrm{g}}m_\mathrm{MP}c^2$ and $\mathbf{p}_\mathrm{MP}=\gamma_{\mathbf{v}_\mathrm{g}}m_\mathrm{MP}\mathbf{v}_\mathrm{g}$, where $\mathbf{v}_\mathrm{g}$ is the group velocity and $\gamma_{\mathbf{v}_\mathrm{g}}=1/\sqrt{1-|\mathbf{v}_\mathrm{g}|^2/c^2}$ is the corresponding Lorentz factor.

\vspace{0.2cm}\noindent\textbf{Gaussian light pulse.}
The electric field of a one-dimensional Gaussian light pulse \cite{Griffiths1998}, linearly polarized in the $y$ direction and propagating in the $x$ direction in vacuum, is given by $\mathbf{E}(\mathbf{r},t)=E_0\cos[k_0(x-ct)]e^{-(\Delta k_0)^2(x-ct)^2/2}\hat{\mathbf{y}}$. Here $\hat{\mathbf{y}}$ is the unit vector in the direction of the $y$ axis, $E_0$ is the electric field amplitude, $k_0=\omega_0/c=2\pi/\lambda_0$ is the wave number, and $\Delta k_0$ is the standard deviation
of the wave number. In all simulations of the present work, we use $E_0=2744.9$ V/m, $\Delta k_0=0.015k_0$, and $\lambda_0=1746.7$ nm. The wavelength is chosen so that it corresponds to the case when the effective phase refractive index of a realistic NIM studied in Ref.~\cite{Valentine2008} becomes $n_\mathrm{p,eff}=-1$. We have selected the NIM structures of the present work so that this condition is also satisfied for them. The magnetic field corresponding to the electric field above is determined by Maxwell's equations. The fields inside materials also follow from Maxwell's equations when the pulse penetrates through the vacuum-material interface. The simulations are carried out using Comsol Multiphysics simulation tool \cite{Comsol2020}. Periodic boundary conditions are applied in the direction of the $y$ axis in Fig.~\ref{fig:inhomogeneoussetup}.

\vspace{0.2cm}\noindent\textbf{Dispersion relation of the EM field.}
In the illustration of light in a homogeneous NIM, we use a dispersion relation linearized around the central angular frequency as
$\omega(k)=\omega_0+(c/n_\mathrm{g})(k-n_\mathrm{p}k_0)$. This dispersion relation corresponds to the frequency-dependent phase refractive index $n_\mathrm{p}(\omega)=n_\mathrm{g}+(n_\mathrm{p}-n_\mathrm{g})\omega_0/\omega$. The group refractive index is constant as $n_\mathrm{g}(\omega)=\frac{\partial}{\partial\omega}[\omega n_\mathrm{p}(\omega)]=n_\mathrm{g}$. This means that the light pulse envelope preserves its shape while the pulse is propagating. To make reflection of light from the vacuum-NIM interface negligible, we match the wave impedance of NIM to that of free space by assuming equal frequency dependencies for the permittivity and permeability by writing $\varepsilon(\omega)=\varepsilon_0n_\mathrm{p}(\omega)$ and $\mu(\omega)=\mu_0n_\mathrm{p}(\omega)$. These are both negative at $\omega_0$ if $n_\mathrm{p}$ is negative. Thus, the condition for negative refraction, i.e., $\varepsilon|\mu|+\mu|\varepsilon|<0$ \cite{Depine2004}, is satisfied in our example. For the phase and group refractive indices at $\lambda_0=1746.7$ nm, we use the values of $n_\mathrm{p}=-1$ and $n_\mathrm{g}=10.52$ corresponding to the central frequency values of the effective refractive indices calculated for the inhomogeneous NIM. The plasmonic host material of the inhomogeneous NIM in Fig.~\ref{fig:inhomogeneoussetup} is assumed to obey the Drude model dispersion relation with a relative permittivity equal to $\varepsilon_\mathrm{h}/\varepsilon_0=1-\frac{\omega_\mathrm{p}^2}{\omega(\omega+i\Gamma)}$. Here the plasma frequency is $\omega_\mathrm{p}=1.22573\times10^{15}$ rad/s. The collision frequency $\Gamma$ is set to zero so that the EM energy is conserved and we can study the field-driven MDW, which is separate from the impulse taken by atoms in absorption. The dielectric cylinders in Fig.~\ref{fig:inhomogeneoussetup} have a constant relative permittivity of $\varepsilon_\mathrm{c}/\varepsilon_0=50.47$, which is the value used for a similar structure in Ref.~\cite{Costa2011}. The plasmonic material used in the schematic experimental setup of Fig.~\ref{fig:setup} has a plasma frequency of $\omega_\mathrm{p}=1.2459\times10^{15}$ rad/s.

\vspace{0.4cm}\noindent\textbf{Mechanical constants of the materials.} In the simulations, we have used the equilibrium mass density of $\rho_\mathrm{a0}=2200$ kg/m$^3$ for the solid materials. The elastic force density was neglected in the present time-dependent simulations since its effect is very small in the short time scale of the propagation of the light pulse \cite{Partanen2018b,Partanen2017c,Partanen2017e}. The elastic force density can be added in Newton's equation of motion without any change in the theory of the coupled field-material dynamics. It is important if one wants to analyze the relaxation of the nonequilibrium atomic displacements produced by the optical force density of the light pulse. This relaxation takes place in the longer time scale of sound waves. In the case of continuous-wave illumination in the experimental setup in Fig.~\ref{fig:setup}, the role of elasticity is to balance the optical force on the layer-NIM by the opposite reaction force resulting from the bending of the cantilever. Thus, in this example, the elastic force density should be added in the simulations if one wants to simulate the bending of the cantilever. In this work, we restrict our simulations of this setup to the calculation of the optical force.

\vspace{0.3cm}\noindent\textbf{DATA AVAILABILITY}\vspace{0.2cm}\\
The simulation results presented in this paper are available from the corresponding author on reasonable request.

\vspace{0.1cm}\noindent\textbf{Acknowledgements}
This work has been funded by the Academy of Finland under Contract No.~318197 and European Union's Horizon 2020 Marie Sk\l{}odowska-Curie Actions (MSCA) individual fellowship under Contract No.~846218. Aalto Science-IT is acknowledged for computational resources.

\vspace{0.1cm}\noindent\textbf{Author Contributions}
M.P. chose the topic, performed the calculations, and wrote the first draft of the manuscript. The manuscript was revised by both authors.


\begin{thebibliography}{10}
\newcommand{\enquote}[1]{``#1''}

\bibitem{ChenJi2011}
J.~Chen, Y.~Wang, B.~Jia, T.~Geng, X.~Li, L.~Feng, W.~Qian, B.~Liang, X.~Zhang,
  M.~Gu, and S.~Zhuang, \enquote{Observation of the inverse {D}oppler effect in
  negative-index materials at optical frequencies,} \emph{Nat. Photon.}
  \textbf{5}, 239 (2011).

\bibitem{Pendry2000}
J.~B. Pendry, \enquote{Negative refraction makes a perfect lens,} \emph{Phys.
  Rev. Lett.} \textbf{85}, 3966 (2000).

\bibitem{Shelby2001}
R.~A. Shelby, D.~R. Smith, and S.~Schultz, \enquote{Experimental verification
  of a negative index of refraction,} \emph{Science} \textbf{292}, 77 (2001).

\bibitem{Shalaev2007}
V.~M. Shalaev, \enquote{Optical negative-index metamaterials,} \emph{Nat.
  Photon.} \textbf{1}, 41 (2007).

\bibitem{Suzuki2018}
T.~Suzuki, M.~Sekiya, T.~Sato, and Y.~Takebayashi, \enquote{Negative refractive
  index metamaterial with high transmission, low reflection, and low loss in
  the terahertz waveband,} \emph{Opt. Express} \textbf{26}, 8314 (2018).

\bibitem{Valentine2008}
J.~Valentine, S.~Zhang, T.~Zentgraf, E.~Ulin-Avila, D.~A. Genov, G.~Bartal, and
  X.~Zhang, \enquote{Three-dimensional optical metamaterial with a negative
  refractive index,} \emph{Nature} \textbf{455}, 376–379 (2008).

\bibitem{Smith2004}
D.~R. Smith, J.~B. Pendry, and M.~C.~K. Wiltshire, \enquote{Metamaterials and
  negative refractive index,} \emph{Science} \textbf{305}, 788 (2004).

\bibitem{Pendry2004}
J.~B. Pendry, \enquote{A chiral route to negative refraction,} \emph{Science}
  \textbf{306}, 1353 (2004).

\bibitem{Kadic2019}
M.~Kadic, G.~W. Milton, M.~van Hecke, and M.~Wegener, \enquote{3{D}
  metamaterials,} \emph{Nat. Rev. Phys.} \textbf{1}, 198 (2019).

\bibitem{Veselago1968}
V.~G. Veselago, \enquote{The electrodynamics of substances with simultaneously
  negative values of $\varepsilon$ and $\mu$,} \emph{Sov. Phys. Uspekhi}
  \textbf{10}, 509 (1968).

\bibitem{Landau1989}
L.~D. Landau and E.~M. Lifshitz, \emph{The Classical Theory of Fields},
  Pergamon, Oxford (1989).

\bibitem{Sun2015}
W.~Sun, S.~B. Wang, J.~Ng, L.~Zhou, and C.~T. Chan, \enquote{Analytic
  derivation of electrostrictive tensors and their application to optical force
  density calculations,} \emph{Phys. Rev. B} \textbf{91}, 235439 (2015).

\bibitem{Wang2016a}
S.~Wang, J.~Ng, M.~Xiao, and C.~T. Chan, \enquote{Electromagnetic stress at the
  boundary: Photon pressure or tension?} \emph{Science Advances} \textbf{2},
  e1501485 (2016).

\bibitem{Wang2016b}
G.~Wang, W.~Zhang, J.~Lu, and H.~Zhao, \enquote{Dispersion and optical gradient
  force from high-order mode coupling between two hyperbolic metamaterial
  waveguides,} \emph{Physics Letters A} \textbf{380}, 2774 (2016).

\bibitem{Rakich2010}
P.~T. Rakich, P.~Davids, and Z.~Wang, \enquote{Tailoring optical forces in
  waveguides through radiation pressure and electrostrictive forces,}
  \emph{Opt. Express} \textbf{18}, 14439 (2010).

\bibitem{Rakich2011}
P.~T. Rakich, Z.~Wang, and P.~Davids, \enquote{Scaling of optical forces in
  dielectric waveguides: {R}igorous connection between radiation pressure and
  dispersion,} \emph{Opt. Lett.} \textbf{36}, 217 (2011).

\bibitem{Pernice2009}
W.~H.~P. Pernice, M.~Li, K.~Y. Fong, and H.~X. Tang, \enquote{Modeling of the
  optical force between propagating lightwaves in parallel 3{D} waveguides,}
  \emph{Opt. Express} \textbf{17}, 16032 (2009).

\bibitem{Costa2011}
J.~T. Costa, M.~G. Silveirinha, and A.~Al\`u, \enquote{Poynting vector in
  negative-index metamaterials,} \emph{Phys. Rev. B} \textbf{83}, 165120
  (2011).

\bibitem{Costa2009}
J.~T. Costa, M.~G. Silveirinha, and S.~I. Maslovski, \enquote{Finite-difference
  frequency-domain method for the extraction of effective parameters of
  metamaterials,} \emph{Phys. Rev. B} \textbf{80}, 235124 (2009).

\bibitem{Silverinha2007}
M.~G. Silveirinha, \enquote{Metamaterial homogenization approach with
  application to the characterization of microstructured composites with
  negative parameters,} \emph{Phys. Rev. B} \textbf{75}, 115104 (2007).

\bibitem{Andren2020}
D.~Andrén, D.~G. Baranov, S.~Jones, G.~Volpe, R.~Verre, and M.~Käll,
  \enquote{Microscopic metavehicles powered and steered by embedded optical
  metasurfaces,} \emph{Nature Nanotechnology} \textbf{16}, 970 (2020).

\bibitem{Partanen2021b}
M.~Partanen and J.~Tulkki, \enquote{Covariant theory of light in a dispersive
  medium,} \emph{Phys. Rev. A} \textbf{104}, 023510 (2021).

\bibitem{Partanen2022a}
M.~Partanen and J.~Tulkki, \enquote{Time-dependent optical force theory for
  optomechanics of dispersive 3{D} photonic materials and devices,}
  \emph{\normalfont{{a}rXiv:2112.11128, Submitted} (2021)} .

\bibitem{Partanen2017c}
M.~Partanen, T.~H\"ayrynen, J.~Oksanen, and J.~Tulkki, \enquote{Photon mass
  drag and the momentum of light in a medium,} \emph{Phys. Rev. A} \textbf{95},
  063850 (2017).

\bibitem{Partanen2017e}
M.~Partanen and J.~Tulkki, \enquote{Mass-polariton theory of light in
  dispersive media,} \emph{Phys. Rev. A} \textbf{96}, 063834 (2017).

\bibitem{Partanen2019a}
M.~Partanen and J.~Tulkki, \enquote{Lorentz covariance of the mass-polariton
  theory of light,} \emph{Phys. Rev. A} \textbf{99}, 033852 (2019).

\bibitem{Partanen2019b}
M.~Partanen and J.~Tulkki, \enquote{Lagrangian dynamics of the coupled
  field-medium state of light,} \emph{New J. Phys.} \textbf{21}, 073062 (2019).

\bibitem{Landau1984}
L.~D. Landau, E.~M. Lifshitz, and L.~P. Pitaevskii, \emph{Electrodynamics of
  Continuous Media}, Pergamon, Oxford (1984).

\bibitem{Jackson1999}
J.~D. Jackson, \emph{Classical Electrodynamics}, Wiley, New York (1999).

\bibitem{Penfield1967}
P.~Penfield and H.~A. Haus, \emph{Electrodynamics of Moving Media}, MIT Press,
  Cambridge, MA (1967).

\bibitem{Ward2005}
D.~W. Ward, K.~A. Nelson, and K.~J. Webb, \enquote{On the physical origins of
  the negative index of refraction,} \emph{New J. Phys.} \textbf{7}, 213
  (2005).

\bibitem{Barnett2015}
S.~M. Barnett and R.~Loudon, \enquote{Theory of radiation pressure on
  magneto{\textendash}dielectric materials,} \emph{New J. Phys.} \textbf{17},
  063027 (2015).

\bibitem{Melcher2014}
J.~Melcher, J.~Stirling, F.~G. Cervantes, J.~R. Pratt, and G.~A. Shaw,
  \enquote{A self-calibrating optomechanical force sensor with femtonewton
  resolution,} \emph{Applied Physics Letters} \textbf{105}, 233109 (2014).

\bibitem{Evans2014}
D.~R. Evans, P.~Tayati, H.~An, P.~K. Lam, V.~S.~J. Craig, and T.~J. Senden,
  \enquote{Laser actuation of cantilevers for picometre amplitude dynamic force
  microscopy,} \emph{Sci. Rep.} \textbf{4}, 5567 (2014).

\bibitem{Kleckner2006}
D.~Kleckner and D.~Bouwmeester, \enquote{Sub-kelvin optical cooling of a
  micromechanical resonator,} \emph{Nature} \textbf{444}, 75 (2006).

\bibitem{Ma2015}
D.~Ma, J.~L. Garrett, and J.~N. Munday, \enquote{Quantitative measurement of
  radiation pressure on a microcantilever in ambient environment,} \emph{Appl.
  Phys. Lett.} \textbf{106}, 091107 (2015).

\bibitem{Ma2018}
D.~Ma and J.~N. Munday, \enquote{Measurement of wavelengthdependent radiation
  pressure from photon reflection and absorption due to thin film
  interference,} \emph{Sci. Rep.} \textbf{8}, 15930 (2018).

\bibitem{Partanen2021a}
M.~Partanen, H.~Lee, and K.~Oh, \enquote{Quantitative in situ measurement of
  optical force along a strand of cleaved silica optical fiber induced by the
  light guided therewithin,} \emph{Photonics Res.} \textbf{9}, 2016 (2021).

\bibitem{Partanen2020b}
M.~Partanen, H.~Lee, and K.~Oh, \enquote{Radiation pressure measurement using a
  macroscopic oscillator in an ambient environment,} \emph{Sci. Rep.}
  \textbf{10}, 20419 (2020).

\bibitem{Vaskuri2021}
A.~K. Vaskuri, D.~W. Rahn, P.~A. Williams, and J.~H. Lehman, \enquote{Absolute
  radiation pressure detector using a diamagnetically levitating test mass,}
  \emph{Optica} \textbf{8}, 1380 (2021).

\bibitem{Astrath2014}
N.~G.~C. Astrath, L.~C. Malacarne, M.~L. Baesso, G.~V.~B. Lukasievicz, and
  S.~E. Bialkowski, \enquote{Unravelling the effects of radiation forces in
  water,} \emph{Nat. Commun.} \textbf{5}, 4363 (2014).

\bibitem{Philbin2011}
T.~G. Philbin, \enquote{Electromagnetic energy momentum in dispersive media,}
  \emph{Phys. Rev. A} \textbf{83}, 013823 (2011).

\bibitem{Jones1978}
R.~V. Jones and B.~Leslie, \enquote{The measurement of optical radiation
  pressure in dispersive media,} \emph{Proc. R. Soc. Lond. A} \textbf{360}, 347
  (1978).

\bibitem{Griffiths1998}
D.~J. Griffiths, \emph{Introduction to Electrodynamics}, Prentice-Hall, Upper
  Saddle River, NJ (1998).

\bibitem{Comsol2020}
\emph{Wave Optics Module User's Guide},
  COMSOL~Multiphysics\textsuperscript{\textregistered}~v.~5.6, COMSOL AB,
  Stockholm, Sweden (2020).

\bibitem{Depine2004}
R.~A. Depine and A.~Lakhtakia, \enquote{A new condition to identify isotropic
  dielectric-magnetic materials displaying negative phase velocity,}
  \emph{Microw. Opt. Technol. Lett.} \textbf{41}, 315 (2004).

\bibitem{Partanen2018b}
M.~Partanen and J.~Tulkki, \enquote{Light-driven mass density wave dynamics in
  optical fibers,} \emph{Opt. Express} \textbf{26}, 22046 (2018).

\end{thebibliography}
\end{document}